\documentclass[sigconf]{acmart}  

\AtBeginDocument{%
  \providecommand\BibTeX{{%
    \normalfont B\kern-0.5em{\scshape i\kern-0.25em b}\kern-0.8em\TeX}}}

\copyrightyear{2022} 
\acmYear{2022} 
\setcopyright{acmlicensed}\acmConference[MSR '22]{19th International Conference on Mining Software Repositories}{May 23--24, 2022}{Pittsburgh, PA, USA}
\acmBooktitle{19th International Conference on Mining Software Repositories (MSR '22), May 23--24, 2022, Pittsburgh, PA, USA}
\acmPrice{15.00}
\acmDOI{10.1145/3524842.3528486}
\acmISBN{978-1-4503-9303-4/22/05}

\usepackage{enumitem}       
\usepackage{hyperref}       
\usepackage{multirow}       
\usepackage{array}          
\usepackage{booktabs}       
\usepackage{makecell}       
\usepackage{tabularx}       
\usepackage{lipsum}         

\newcommand{\mytilde}{\raise.17ex\hbox{$\scriptstyle\mathtt{\sim}$}}  

\newcommand\inc{Incorporate}

\begin{document}

\title{An Alternative Issue Tracking Dataset of Public Jira Repositories}


\author{Lloyd Montgomery}
\email{lloyd.montgomery@uni-hamburg.de}
\orcid{0000-0002-8249-1418}
\affiliation{%
  \institution{University of Hamburg}
  \city{Hamburg}
  \country{Germany}
}

\author{Clara L{\"u}ders}
\email{clara.marie.lueders@uni-hamburg.de}
\orcid{0000-0001-7743-4067}
\affiliation{%
  \institution{University of Hamburg}
  \city{Hamburg}
  \country{Germany}
}

\author{Walid Maalej}
\email{walid.maalej@uni-hamburg.de}
\orcid{0000-0002-6899-4393}
\affiliation{%
  \institution{University of Hamburg}
  \city{Hamburg}
  \country{Germany}
}


\begin{abstract}
Organisations use issue tracking systems (ITSs) to track and document their projects' work in units called issues.
This style of documentation encourages evolutionary refinement, as each issue can be independently improved, commented on, linked to other issues, and progressed through the organisational workflow.
Commonly studied  ITSs so far include GitHub, GitLab, and Bugzilla, while Jira, one of the most popular ITS in practice with a wealth of additional information, has yet to receive similar attention.
Unfortunately, diverse public Jira datasets are rare, likely due to the difficulty in finding and accessing these repositories.
With this paper, we release a dataset of 16 public Jiras with 1822 projects, spanning 2.7 million issues with a combined total of 32 million changes, 9 million comments, and 1 million issue links.
We believe this Jira dataset will lead to many fruitful research projects investigating issue evolution, issue linking, cross-project analysis, as well as cross-tool analysis when combined with existing well-studied ITS datasets.
\end{abstract}




\maketitle

\section{Introduction}




\begin{table}[b]
\small
\setlength\tabcolsep{1pt}  
\centering
\captionsetup{skip=0pt}
\caption{Dataset consisting of 16 public Jira repositories}
\label{table:datasource}
\begin{tabular}{lrrrrrrrrrr}
\multicolumn{11}{c}{\footnotesize{\makecell{
    Column names: Documented Issue Types (DIT); Used Issue Types (UIT); \\
    Documented Link Types (DLT); Used Link Types (ULT); \\
    Changes per Issue (Ch/I); Comments per Issue (Co/I); Unique Projects (UP)}}} \\
\toprule
Jira repo     & Born & Issues    & DIT & UIT & Links   & DLT & ULT & Ch/I & Co/I & UP   \\
\midrule
Apache        & 2000 & 1,014,926 & 48  & 49  & 264,108 & 20  & 20  & 10   & 5    & 657  \\
Hyperledger   & 2016 & 28,146    & 9   & 9   & 16,846  & 6   & 6   & 12   & 2    & 36   \\
IntelDAOS     & 2016 & 9,474     & 4   & 11  & 2,667   & 12  & 12  & 15   & 3    & 7    \\
JFrog         & 2006 & 15,535    & 30  & 22  & 3,303   & 17  & 10  & 9    & 1    & 35   \\
JIRA*         & 2002 & 274,545   & 52  & 38  & 110,507 & 19  & 18  & 21   & 3    & 123  \\
JiraEcosystem & 2004 & 41,866    & 122 & 40  & 12,439  & 19  & 18  & 15   & 2    & 153  \\
MariaDB       & 2009 & 31,229    & 11  & 10  & 14,950  & N/A & 6   & 12   & N/A  & 24   \\
Mindville     & 2015 & 2,134     & 2   & 2   & 46      & N/A & 4   & 3    & N/A  & 10   \\
Mojang        & 2012 & 420,819   & 1   & 3   & 215,821 & 6   & 5   & 8    & 2    & 16   \\
MongoDB       & 2009 & 137,172   & 31  & 37  & 92,368  & 15  & 13  & 17   & 3    & 95   \\
Qt            & 2005 & 148,579   & 19  & 14  & 41,426  & 10  & 10  & 12   & 3    & 38   \\
RedHat        & 2001 & 353,000   & 74  & 55  & 163,085 & 23  & 19  & 13   & 2    & 472  \\
Sakai         & 2004 & 50,550    & 43  & 15  & 20,292  & 7   & 7   & 10   & 4    & 55   \\
SecondLife    & 2007 & 1,867     & 10  & 12  & 674     & 9   & 5   & 30   & 8    & 3    \\
Sonatype      & 2008 & 87,284    & 16  & 21  & 4,975   & 10  & 9   & 7    & 4    & 17   \\
Spring        & 2003 & 69,156    & 13  & 14  & 14,716  & 7   & 7   & 8    & 3    & 81   \\
\midrule
Sum           &      & 2,686,282 & 485 & 352 & 978,223 & 180 & 169 &      &      & 1822 \\
Median        &      & 59,853    & 18  & 14  & 15,898  & 10  & 10  & 12   & 3    & 37   \\
Std Dev       &      & 251,973   & 31  & 16  & 81,744  & 6   & 5   & 6    & 2    & 178  \\
\bottomrule
\multicolumn{11}{l}{\footnotesize{\makecell[l]{* Note that our data source tool name is ``Jira'' while the organisation ``JIRA'' is\\written in all capitals to help distinguish the two throughout the article.}}} \\
\end{tabular}
\end{table}

Software projects use Issue Tracking Systems (ITSs) to collect and track issues that need to be addressed such as bug reports and feature requests.
Issues are units of information usually including  a summary (title), a description, and a number of properties like status, priority, and fix version.
A key focus of ITSs is the evolutionary refinement of the issues~\cite{Ernst_empiRE_2012} (also known as iterative improvement), which means that information is gained and refined over time, while developers and stakeholders collaborate to address the issues.

Over the last two decades, software engineering research has intensively studied issues and issue trackers, often based on Bugzilla\footnote{\url{https://www.bugzilla.org}} and GitHub\footnote{\url{https://github.com}}.
The primary research focus has been on the specific issue type of \textit{bug reports}:  
the understanding and improvement of information quality therein~\cite{Bettenburg_FSE_2008,Zimmermann_TSE_2010} and the prediction of bug properties 
such as severity~\cite{Lamkanfi_MSR_2010, Lamkanfi_CSMR_2011}, assignee~\cite{Jeong_FSE_2009}, and duplicate reports~\cite{Wang_ICSE_08,Deshmukh_ICSME_2017,He_ICPC_2020} for supporting software evolution and maintenance. 

Another prominent but rather understudied issue tracker is Jira.
Jira\footnote{\url{https://www.atlassian.com/software/jira}} is an agile planning platform that offers features such as scrum boards, kanban boards, and roadmap management, beyond issue tracking.
Jira's ticket-centric design mimics that of GitHub\footnote{\url{https://docs.github.com/en/issues}}, GitLab\footnote{\url{https://docs.gitlab.com/ee/user/project/issues/}}, and Bugzilla.
Benefits of Jira over other ITSs includes a history of issue changes, complex issue linking networks, and a diverse set of custom field configurations across organisations.
According to slintel\footnote{\url{www.slintel.com/tech/bug-and-issue-tracking/atlassian-jira-market-share}}, 
Datalyze\footnote{\url{www.datanyze.com/market-share/project-management--217/jira-market-share}}, 
and Enlyft\footnote{\url{https://enlyft.com/tech/products/atlassian-jira}}, Jira is by far the most popular tool in the issue tracking and agile project management markets.
Yet it is rather under-represented in software engineering research: Google Scholar article title search: ``GitHub'' returns 4,450 results, ``Jira'' returns 650.
We believe the reason for such a lack of wide-spread research on Jira is the lack of available and diverse Jira data to study.
While Ortu et al. released a dataset of four Jira repos in 2015\cite{Ortu_PROMISE_2015}, our dataset is larger and more diverse.

We collected data from 16 pubic Jira repositories (repos) containing 1822 projects and 2.7 million issues.
Our dataset includes historical records of 32 million changes, 9 million comments, as well as 1 million issue links that connect the issues in multiple ways.
Table~\ref{table:datasource} shows an overview of the data.
The table lists for each Jira repo the year it first was used, the number of issues, number of documented and used issue types, number of documented and used link types, number of changes per issue, number of comments per issue, and number of unique projects.
Apache is the largest repo with 1 million issues and 10.5 million changes.
Mindville and SecondLife are the smallest repos with \mytilde2,000 issues each.
The number of unique projects is particularly important because each Jira repo contains documentation for multiple projects, e.g.~657 projects for Apache.
While our dataset has 16 different repositories, it represents 1822 actual projects that can be studied.
Our data and code is publicly available with an open source license.\footnote{\url{https://doi.org/10.5281/zenodo.5882881}}

Each Jira repo has a number of documented issue types, but only a subset of them are actually used in the projects. 
As shown in Table~\ref{table:datasource}, 485 types are defined while only 352 are used.
Notably, the median number of used issue types is 14, which is significantly higher than expected from research discussing ITSs (bug reports~\cite{Bettenburg_FSE_2008,Zimmermann_TSE_2010}, feature requests~\cite{Fitzgerald_RE_2011,Merten_RE_2016}, tasks~\cite{Ernst_empiRE_2012}, or technical debt~\cite{Bavota_MSR_2016,xavier_MSR_2020}).
The projects also vary in terms of the number of documented and used link types, totally 186 and 169, respectively.
The median number of used link types is 10 which is considerably more than the mainly well-researched ``duplicate'', or ``depends/relates'' types.

\begin{figure}[t]
    \centering
    \includegraphics[width=\columnwidth]{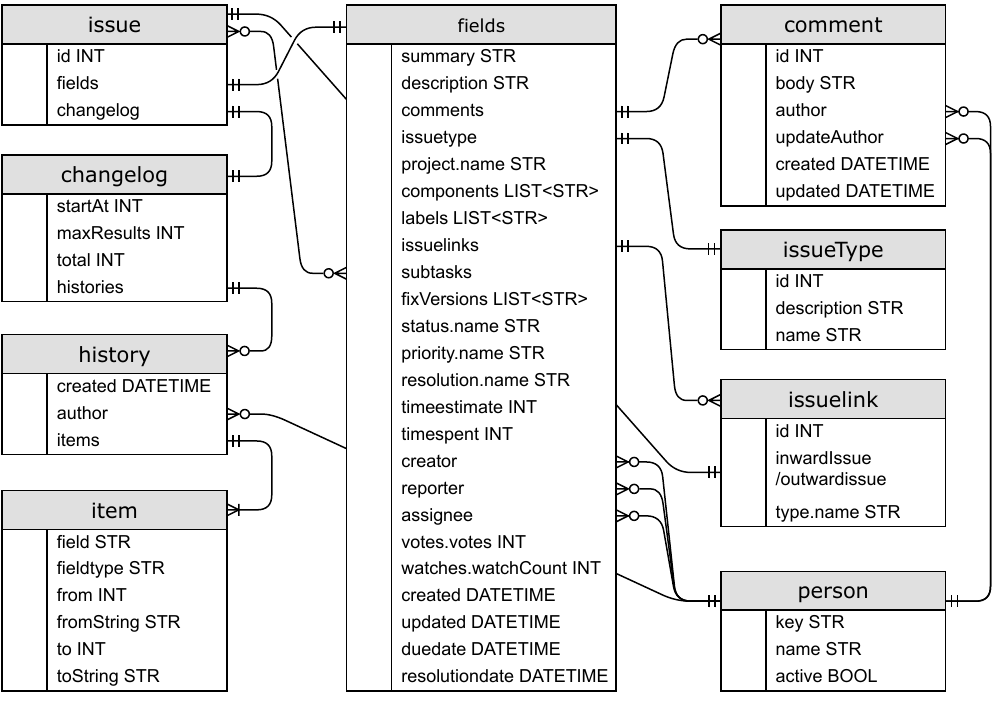}
    \caption{Jira MongoDB database scheme}
    \label{fig:database_erd}
\end{figure}

\section{Dataset Description}

The Jira data is stored in MongoDB as document data.
Figure~\ref{fig:database_erd} describes the data as an entity relationship diagram (ERD).
It is not possible to create a full and reliable ERD for a document database such as MongoDB, unless a strict structure is enforced on all documents, and that is not the case for Jira across different repositories.
Figure~\ref{fig:database_erd}, therefore, is a simplification of the real data structure.
We have diagrammed the key objects in the data
and extrapolated the data types to give an idea of what to expect.
The primary documents to expect in our dataset are ``issues'', each unit of which represents a single issue within Jira.
The two primary documents nested within each issue are ``fields'' and ``changelog''.
The fields document contains the current attributes of the issue, and is described in Table~\ref{table:issue_fields}.
The issue field descriptions are primarily from the Jira official documentation.\footnote{\url{https://confluence.atlassian.com/adminjiraserver/issue-fields-and-statuses-938847116.html}}
The ``changelog'' stores the changes that have occurred to this issue.
When an issue is changed, a new ``history'' is saved, where each changed attribute is stored as an ``item''.
``Issuelinks'' connect issues to each other and are stored on both linked issues.
``Comments'' are made by the community and form a discussion around the issue.

\begin{table}[t]
\centering
\small
\setlength\tabcolsep{2.5pt}  
\renewcommand{\arraystretch}{.9}  
\captionsetup{justification=centering,skip=0pt}  
\caption{Issue fields}
\label{table:issue_fields}
\begin{tabularx}{\columnwidth}{l X c}
\toprule
\textbf{Issue Field} & \textbf{Description}  \\ 
\midrule
& \makecell[c]{\footnotesize\textbf{Content}}  \\
    summary             & A brief one-line summary of the issue.  \\
    description         & A detailed description of the issue.  \\
    comments            & Community discussion on each issue.  \\
& \makecell[c]{\footnotesize\textbf{Context \& Links}}  \\
    issuetype           & The issue purpose within the organisation.  \\
    project             & The parent project to which the issue belongs. \\
    components          & Project component(s) to which this issue relates.   \\
    labels              & Labels to which this issue relates.  \\
    issuelinks          & A list of links to related issues.  \\
    subtasks            & Sub-issues to this issue; can only be one level deep.  \\
& \makecell[c]{\footnotesize\textbf{Workflow}}  \\
    fixVersions         & Project version in which the issue was (or will be) fixed.  \\
    status              & The stage the issue is currently at in its lifecycle.  \\
    priority            & The issue importance in relation to other issues.  \\
    resolution          & A record of the issue's resolution, once resolved or closed.  \\
    timeestimate        & Estimated amount of time required to resolve the issue.  \\
    timespent           & Amount of time spent working on this issue.  \\
& \makecell[c]{\footnotesize\textbf{People \& Community}}  \\
    creator             & The person who created the issue. \\
    reporter            & The person who found/reported the issue. Defaults to the creator unless otherwise assigned. \\
    assignee            &  The person responsible to resolve the issue. \\
    votes               & Number of people who want this issue addressed. \\
    watches             & Number of people watching this issue.  \\
& \makecell[c]{\footnotesize\textbf{Timestamps}}  \\
    created             & Time and date this issue was entered into Jira.  \\
    updated             & Time and date this issue was last edited.  \\
    resolutionedate     & Time and date this issue was resolved.  \\
    duedate             & Time and date this issue is scheduled to be completed.  \\
\bottomrule
\end{tabularx}
\end{table}


The dataset is available on Zenodo as a MongoDB dump file.
The raw data includes a README file, the scripts used to download the data, the scripts used to produce the tables and figures in this paper, and a license (CC BY 4.0\footnote{\url{https://creativecommons.org/licenses/by/4.0/}}).
In principle, using this dataset is as easy as importing the MongoDB dump file and modifying the included JupyterLab notebook to explore the data.
The data can also be queried using standard MongoDB queries.
Figure~\ref{fig:database_erd} can be used to guide the initial exploration.


The data was downloaded using a Python script utilising the Jira API exposed by public Jira ITSs.
We obtained the list of repos through a manual search for public Jira repos on the internet.
This search involved reviewing GoogleScholar results for ``Jira'', as well as Google search results discussing a public Jira repo.
With the public Jira URLs, the data was downloaded using the Jira REST API V2\footnote{\url{https://developer.atlassian.com/cloud/jira/platform/rest/v2}} and a Python script contained within a JupyterLab notebook.
In total, \mytilde15GB of data was downloaded and stored in the MongoDB database.
The initial data download was performed in May 2021 and the data was updated in January 2022.
Unfortunately, MariaDB and Mindville are no longer valid data sources, so their data is from the initial download in May 2021.


\section{Enhancing the Dataset}

To enrich the dataset and support the comparison of the repositories, we enhanced the dataset through the addition of manual qualitative labelling of the issue types and link types.

\subsection{Issue Type Mapping}


Jira is a highly customisable tool, where each organisations defines specific aspects such the issue types they use.
The issue type indicates the general organisational, workflow category the issue belongs to, and likely has impact on the organisational workflow and subsequent research analysis.
Given the 352 unique issue types (see Table~\ref{table:datasource}), it is unclear which issue types are offered by each Jira, how to group them, and then how to unify them across the Jiras for more general analyses.
To address this, we mapped the unique issue types to software life-cycle activities they represent.

We conducted a thematic analysis~\cite{Braun_QRP_2006,Cruzes_ESEM_2011} of all issue types across all 16 Jira repos to provide a unified mapping between the issue types.\footnote{The thematic analysis was performed using the May 2021 data.}
We followed the recommendations by Braun and Clarke~\cite{Braun_QRP_2006} and Cruzes and Dyba~\cite{Cruzes_ESEM_2011}, including 5 main analysis steps.
First, we familiarised ourselves with the data by looking at the issue type descriptions in the repos and 50 example issues for each used type.
Second, we merged issue types with similar names and meaning, and identified initial phrases or codes describing the types.
Third, we iterated through the codes and example issues and identified an initial set of themes.
Fourth, we reviewed and reduced the themes during two peer-discussion sessions.
Finally, we finalised our themes by forming definitions for each.

Our thematic analysis produced a mapping between each issue type in each Jira and a unified set of themes and codes.\footnote{The codes are available in our replication package and are not discussed here.}
The themes (and issue proportions) are Requirements (30\%), Development (18\%), Maintenance (51\%), User Support (1\%), and Other (<1\%).
This mapping (stored as a JSON) allows one to programmatically ask questions such as ``give me the `Requirements' issues for Apache''.

In the following we describe the five themes in more detail.
\textbf{Requirements} activities are documented in ITSs through lightweight representations~\cite{Ernst_empiRE_2012} such as epics, user stories, and feature requests.
Both top-down as well as bottom-up requirements are captured in the ITSs.
We found top-down epics and stories, as well as bottom-up feature requests and change requests.
This shows the breadth of requirements knowledge maintained within ITSs.
\textbf{Development} issue types track development activities such as what needs to be done and who is doing it.
Example issue types include Task, Technical Task, Dev Task, and Sub-Task.
\textbf{Maintenance} issue types correspond to maintenance activities documented within ITSs through both bottom-up (Bug, Defect, Incident) as well as top-down (Technical Debt, Documentation, QE Task) work.
This includes quality assurance issues, legacy upgrades, and continuous integration activities.
\textbf{User Support} issues directly assist users of software systems with respect to how to use the product.
While users can submit Bug Reports to the Maintenance theme above, Bug Reports focus on \textit{issues with the code or product}, while issues submitted to the User Support theme are focused on \textit{the user and their use or interpretation of the code or product}.
Example issue types include Support Request, Problem Ticket, IT Help, and Question.
\textbf{Other} issue types are not representative across the Jira repos, or could not be categorised at all.
Only two Jira repos contained organisational process issues, and many issue types were lacking enough examples to understand what they were documenting.
Example issue types include New Project, GitHub Integration, Fug, and Spike.

\begin{table}[b]
\centering
\small
\setlength\tabcolsep{2pt}  
\renewcommand{\arraystretch}{.9}  
\captionsetup{justification=centering,skip=0pt}  
\caption{Overview of cleaned link types, their description and number of Jiras with that link type.}
\label{tab:LinkType}
\begin{tabularx}{\columnwidth}{l X c}
\toprule
\textbf{Link Type} & \textbf{Description} & \textbf{Jiras}  \\ 
\midrule
Relate & Two issues are related to each other in a general way or stakeholders are unsure about the nature of the connection. & 16 \\
Duplicate & An issue is already in the system and someone added the same issue by accident. & 15 \\
Subtask & An issue is a part of a larger set of issues. & 14 \\
Clone & An issue was duplicated intentionally by the users, e.g., when an old bug resurfaces in a newer version. & 14 \\
Block & An issue cannot be resolved until another issue is resolved. & 11 \\
Depend & An issue cannot be resolved until another issue is resolved. & 10 \\
Epic & A link that consists of higher-level issue which has multiple low-level issues attached to it, similar to Parent-Child or Incorporate.  & 9 \\
Incorporate & An issue is included or contained in another issue. E.g., an issue is used to collect multiple other issues, similar to Parent-Child. & 9 \\
Split & An issue is split into multiple issues to reduce the size or complexity. & 9 \\
Cause & Issue A describes a problem that is at least a partial root cause for Issue B. & 7 \\
Bonfire Testing & An issue was discovered while testing another issue. & 7 \\
finish-finish & Two issues must be finished jointly. & 4 \\
start-start & Two issues must be started jointly. & 4 \\
Supercede & An issue was refined based on existing issues.  & 4 \\
\bottomrule
\end{tabularx}
\end{table}

\subsection{Link Type Mapping}

One of the main strengths of Jira consists of its ability to link issues and form complex issue graphs.
An ``issue link'' denotes a relationship between two issues including a name and a description.
Examples of issue links include ``duplicate'', ``clone'', and ``depend.''
Issue links are an important part of Jira and issue trackers in general.
More research is needed to understand how issue linking works, what effect links have on the issue tracking ecosystems, and how they can be leveraged to create more productive and efficient development environments.
While the ``duplicate'' issue links have been studied in great detail, our dataset reveals that there are obviously many more to be explored.
One potential problem is the sheer number of link types that exist and the lack of uniformity across different Jira repos.
To address this problem, we performed a qualitative labelling process similarly to the issue types to unify the 90 uniquely named link types across the 16 Jira repos.

To create a unified set of link types, we first grouped the link types together based on types with similar word stems (e.g. ``finish-finish [gantt]'', ``gantt: finish-finish'' or ``Depend'', ``Dependency'', ``Dependent'', ``Depends''). 
Then, we compared the link types by their meaning.
We looked up random examples in the respective ITS and reviewed the texts and properties of the involved issues as well as the language used to describe the link.
For instance, we saw that 9 projects had a different link type name for ``\inc{}'', but all 8 used ``includes'' or synonyms such as ``contains'' or ``incorporates''.
After this step, we arrived at 34 distinct link types.
We list the link types that appeared in at least 25\% of the repos in Table~\ref{tab:LinkType} with brief descriptions and the count of repos where they are used.

\section{Relevance of the Dataset}



Studying developers behaviour, organisational processes, and user-developer interactions are common research areas, with work published at large venues such as Mining Software Repositories (MSR), the International Conference on Software Engineering (ICSE), and the International Conference on Requirements Engineering (RE), as well as journals such as Transactions on Software Engineering (TSE), Empirical Software Engineering (EMSE) and Requirements Engineering (REJ).
The breadth and depth of studies that are conducted rely on access to real developers, real organisations, and real data.
Our dataset contributes to this research line of studying real organisational data: including a diversity of issues, issue types, and link types across multiple organisations, developers, and projects.

\paragraph*{Triangulate results from studies on Bugzilla and GitHub Issues.}
There are many studies investigating issues in Bugzilla and GitHub, but very little work done to verify these findings across organisations and platforms.
Our dataset can act as an additional research source to triangulate the findings of those past studies, adding further validity to issue tracking research, such as ``what makes a good bug report''~\cite{Bettenburg_FSE_2008} and ``good first issues for newcomers''~\cite{Steinmacher_IEEES_2018}.
Our dataset is not only fairly large and heterogeneous, but also brings an additional ITS which is popular among practitioners. 

\paragraph*{Issue and link types comparative analyses.}
Issue tracking systems contain much more than just the heavily studied bug reports~\cite{Bettenburg_FSE_2008,Zimmermann_TSE_2010} and feature requests~\cite{Heck_IWPSE_2013}.
By our preliminary analysis, there are at least four major categories: requirements, maintenance, development, and user support.
As a starting place, trying to reliably separate these issue types (beyond just the stated issue type) is a much needed area of research.
Herzig et al.~investigated issues labelled as bug reports and found that roughly 1/3rd are misclassified, leading to the conclusion that data needs to be better classified~\cite{Herzig_ICSE_2013}.
Additionally, studying the differences between these issue types, across organisations and tools, is likely to reveal critical workflow differences key to running successful software projects.
Similarly, the link type ``duplicate'' has been studied in great detail, while the full ecosystem of other link types largely remains unexplored.

\paragraph{Issue evolution studies and support.}
A key focus of ITSs (over traditional requirements) is the ongoing evolutionary refinement of the issues~\cite{Ernst_empiRE_2012}, where information is gained and refined over time as developers and stakeholders collaborate to address the issues.
This evolution has not been studied directly, possibly because no such dataset existed.\footnote{Heck and Zaidman~\cite{Heck_IWPSE_2013} investigated the requirements evolution in ITSs, but to the best of our knowledge did not investigate the historical changes.}
This Jira dataset provides the historical changes for each issue, allowing for a direct investigation of issue evolution.

\paragraph{User support analysis.}
Understanding and supporting users is an important area of SE research.
While bug reports have been thoroughly explored, and feature requests are now receiving much more attention, the concept of customer relationship management is a much broader topic.
Customer relationship management involves integrating artefacts, tools, and workflows to successfully initiate, maintain, and (if necessary) terminate customer relationships~\cite{Reinartz_JMR_2004}.
Jira stores a number of user support issue types, all of which combine together to tell a story about the users of these software products.
Possible areas of investigation include support ticket escalation prediction~\cite{Montgomery_RE_2017} and investigating sentiment of customer and developer communication~\cite{Ortu_MSR_2016}.

\section{Challenges and Limitations}


There were a number of challenges associated with creating this Jira dataset.
\textbf{1.} Unlike GitHub and GitLab, Jira does not have a central public list of repos.
The software is designed to be hosted by individual organisations, who largely opt for closed-access.
For those who do opt for public, there is no central place to find all of these repos, making them very difficult to find.
\textbf{2.} While GitHub, GitLab, and Bugzilla are customisable tools, Jira offers far more options for the customisation of issue types, link types, workflows, automated scripts, custom fields, and linking.
The data is richer, but also more complex.
We have made a first attempt at unifying the complexity by starting with the issue and link types.
\textbf{3.} Downloading this data took weeks, and updating the data presented inconsistencies such as Jira repos no longer existing (MariaDB and Mindville).
\textbf{4.} We present the whole unfiltered data and only made suggestion on how this data set could be used. 
Filtering is an additional work step a researcher must do depending on their research question (e.g. removing all issues that have not yet been resolved).

\section{Summary}
We provide a dataset of public Jira repos with the scripts to re-download the data, the data itself, scripts for an initial investigation, and qualitative labelling to enhance the research potential.
In total, there are 16 Jira repos, with 1822 projects, consisting of 2.7 million issues, with 32 million changes, 9 million comments, and 1 million issue links.
We believe this data, in combination with our qualitative labelling enhancements, serves as an alternative data source for investigating issue tracking in software engineering practice.

\begin{acks}
We thank Abir Bouraffa for help with the issue link labelling.
This work was partly funded by the European Union Horizon 2020 Research and Innovation programme under grant agreement No 732463.
We acknowledge the support of the Natural Sciences and Engineering Research Council of Canada (NSERC), [funding reference number PGSD3-518105-2018].
\end{acks}

\clearpage
\newpage
\balance
\bibliographystyle{ACM-Reference-Format}
\bibliography{main}

\end{document}